\documentclass[%
 reprint,
showpacs,preprintnumbers,
 amsmath,amssymb,
 aps,
 prd,
floatfix,
]{revtex4-1}

\usepackage{xcolor}
\usepackage{graphicx}
\usepackage{dcolumn}
\usepackage{bm}
\usepackage{verbatim}
\usepackage[caption=false]{subfig}

\graphicspath{{images/}}

\begin{document}

\preprint{APS/123-QED}

\title{Prospects for Neutrino Spin Coherence in Supernovae}

\author{James Y. Tian}
 \email{j2tian@ucsd.edu}
 \author{Amol V. Patwardhan}
 \email{apatward@ucsd.edu}
\author{George M. Fuller}%
 \email{gfuller@ucsd.edu}
\affiliation{%
Department of Physics, University of California, San Diego, La
Jolla, California 92093, USA
}%




\date{\today}

\begin{abstract}
We present neutrino bulb model simulations of Majorana neutrino coherent spin transformation (i.e., neutrino-antineutrino transformation), coupled to neutrino flavor evolution, for conditions corresponding to the neutronization burst epoch of an Oxygen-Neon-Magnesium (O-Ne-Mg) core collapse supernova. Significant neutrino spin transformation in, for example, the neutronization burst, could alter the fluences of neutrinos and antineutrinos in a way which is potentially detectable for a Galactic core collapse supernova. Our calculations for the first time incorporate geometric dilution in the spin evolution of the neutrinos and combine two-flavor and three-flavor evolution with spin mixing physics. We find that significant spin transformations can occur, but only with a large neutrino luminosity and an electron fraction ($Y_e$) profile which facilitates adiabatic conditions for the spin-channel resonance. Using our adopted parameters of neutrino energy spectra, luminosity, density and $Y_e$ profiles, our calculations require an unrealistically large neutrino rest mass to sustain the spin transformation. It is an open question whether examining different density profiles or incorporating other sources of nonlinear feedback, such as $Y_e$ feedback, could mitigate this need. We find that spin transformations are not sensitive to the flavor structure of neutrinos, i.e., the spin transformations occur regardless of whether we simulate two- or three-flavor transformations. In the two-flavor case, spin transformations were insensitive to the choice of solar or atmospheric mass-squared splitting as well as the choice of the Majorana phase. Importantly, our three-flavor simulations, as well as our two-flavor simulations done with the atmospheric mass-squared splitting, show that the inclusion of spin degrees of freedom can significantly and qualitatively alter neutrino flavor evolution. 
 
\end{abstract}

\pacs{14.60.Pq, 97.60.Bw, 13.15.+g, 26.30.-k, 26.50.+x}
\maketitle


\section{Introduction}

In this paper we study new aspects of how neutrino flavor and spin physics could play out in the core collapse supernova environment. Neutrino flavor transformation in astrophysical environments can be a complicated, nonlinear phenomenon \cite{Duan06a,Duan06b,Duan06c,Duan07a,Duan07b,Duan07c,Duan08,Duan:2008qy,Duan:2008eb,Cherry:2010lr,Duan:2010fr,Duan:2011fk, Cherry:2011bg, Cherry:2012lu, 2013PhRvD..88j5009Z,2014AIPC.1594..313B,2015AIPC.1666g0001L,2016AIPC.1743d0001B, 2016NuPhB.908..382A,2016JCAP...01..028C,Barbieri:1991fj,2016arXiv160906747V,Balantekin:2007kx, 2016PhRvD..94h3505J, Raffelt:2013qy,Sarikas:2012fk,Raffelt07,Hannestad06,Dasgupta09,Notzold:1988fv, Pastor:2002zl, Dasgupta:2008kx,Cherry:2013lr, Cherry:2012lr, 2015PhRvD..92l5030D, 2015PhLB..751...43A, 2015PhLB..747..139D, 2015IJMPE..2441008D,2015PhRvD..92f5019A, 2014JCAP...10..084D, Qian93,1992AAS...181.8907Q, Qian95, 1995PhRvD..52..656Q,2011PhRvD..84e3013B}. In addition, there have been several studies of neutrino \textit{spin} (or helicity) transformation as a consequence of an external magnetic field acting on a large neutrino magnetic moment, some of which are in the context of supernovae \cite{de-Gouvea:2012fk,de-Gouvea:2013lr,1988PhLB..213...64A,1992NuPhB.373..479A,1997PhRvD..55..515A,1989PhRvL..63..228B, 1990PhRvL..65.2626B,1988PhRvD..37.1368L,1981PhRvD..24.1883S,2015PhRvD..92l5020V,2009PhRvD..80l5032Y,2009PhRvD..79k3015D, 2007arXiv0711.3237V,2007JCAP...09..016B,2003JCAP...12..007A,2003JCAP...11..004A,2003PhRvD..68b3003A, 2003PhRvD..67b3004A,2003JCAP...10..001A,1999APh....11..317N,1997PhRvD..55.3265N,1999PhR...320..319R,1997PhRvD..56.6117A, 2014PhRvD..90f5011P,2015RvMP...87..531G,2013arXiv1310.4340D,2016AnP...528..198G,1992NuPhB.373..479A}. However, it has been discovered recently, via examination of the quantum kinetic equations (QKEs), that neutrinos may undergo this spin conversion from left-handed helicity states to right-handed helicity states purely kinetically in the presence of an asymmetric matter and neutrino flow (as would be present in a supernova environment), even in the absence of a magnetic field or a large magnetic moment \cite{2014arXiv1406.6724V,Volpe:2013lr,Vlasenko:2014lr,2014PhRvD..90l5040S,2015PhyA..432..108A,2015IJMPE..2441009V, 2015PhLB..747...27C,2015PhRvD..91l5020K,2016PhRvD..93l5030D,2016arXiv161101862C,2016JPhCS.718f2068V, 2016PhRvD..94c3009B,Sigl:1993fr}. In this paper we study spin conversions arising from purely kinetic effects.

In vacuum, active neutrinos are in left-handed helicity states and active antineutrinos are in right-handed helicity states. If neutrinos are Majorana in nature, spin transformations are equivalent to transformations of neutrinos into antineutrinos and vice versa. If neutrinos are Dirac in nature, this spin transformation would produce sterile states from the active neutrino species. In this paper, we assume neutrinos are Majorana in nature and examine the prospects for coherent neutrino-antineutrino transformation during the neutronization burst epoch of an O-Ne-Mg core collapse supernova.

In medium, the propagation states of neutrinos can be superpositions of left-handed and right-handed helicity states. As was first shown in Ref. \cite{2014arXiv1406.6724V}, it is possible to find a \lq\lq resonance\rq\rq, akin to a Mikheyev-Smirnov-Wolfenstein (MSW) resonance~\cite{Wolfenstein78,Mikheyev85}, through which adiabatic propagation gives nearly complete helicity flip. However, this spin resonance is narrow, i.e., the instantaneous neutrino energy eigenstates are nearly degenerate through resonance, implying that achieving the conditions required for adiabatic spin transformation is problematic. An outstanding question is whether nonlinear feedback from spin transformation can augment the adiabaticity in a core collapse supernova environment. In this paper we investigate this issue, with the new features here being coupled spin and flavor evolution and a more realistic geometry. 

In seeking the optimal environment for neutrino spin degrees of freedom to affect neutrino evolution, we focus on the core collapse supernova neutronization burst. As a massive star reaches the end of its life, its core becomes dynamically unstable. If the core of the star is sufficiently massive, i.e., over the Chandrasekhar limit, electron degeneracy pressure is overcome by gravity and the core will catastrophically collapse until it reaches nuclear densities \cite{2012PTEP.2012aA309J}. As the core collapses, it \lq\lq neutronizes\rq\rq\ via charged current electron capture on protons in heavy nuclei. An inner, homologous, core \lq\lq bounces\rq\rq\ at nuclear density and serves as a piston, driving a shock into the outer part of the core \cite{2012PTEP.2012aA309J,2015arXiv150101688M,2014ASPC..488..102M}. When this shock comes through the \lq\lq neutrino sphere\rq\rq\ (roughly coincident with the outer edge of the core), where the material becomes more or less transparent to neutrinos, we get a \lq\lq neutronization burst\rq\rq\ \cite{2015AIPC.1666g0001L,2016NCimR..39....1M}. This shock breakout, lasting $\approx 10\,\text{ms}$, is accompanied by a spike in the neutrino luminosity of order $10^{53}$ to $10^{54}$ $\text{erg\,s}^{-1}$. Moreover, the flavor content of this neutronization burst is overwhelmingly electron type neutrinos, $\nu_e$ \cite{2016NCimR..39....1M}.   

In this paper we examine the prospects for neutrino spin transformations specifically in the neutronization burst epoch for two main reasons. First, since the neutronization burst neutrino luminosities are extremely high \cite{2016NCimR..39....1M}, there can be a larger contribution to the $\nu_e \rightleftharpoons \bar{\nu}_e$ transformation channel in the Hamiltonian during the neutronization burst than during other epochs. This may lead to conditions which are the most favorable for coherent spin transformation. Second, since the neutronization burst produces an overabundance of electron neutrinos over all other flavor and spin states \cite{2016NCimR..39....1M}, spin transformations, if they occur, can drastically change the ratio of left-handed neutrinos to right-handed antineutrinos coming out of the supernova. This therefore makes spin transformations during the neutronization burst a potentially measurable event. Detection of a neutronization burst in a terrestrial detector, e.g., the Deep Underground Neutrino Experiment (DUNE) or Hyper-Kamiokande (Hyper-K), could provide, in principle, a unique way to probe neutrino absolute masses and Majorana phases complementary to neutrinoless double beta decay experiments. In a hypothetical example, suppose Hyper-K detects a significant antineutrino content in the neutronization burst of a future galactic core-collapse supernova. What would that imply for parameters such as the neutrino absolute rest-mass scale, or matter density and electron fraction profiles in the envelope? What would that mean for models of neutrino heating or nucleosynthesis? Answering these questions requires detailed calculations.
   
For this paper, we conducted astrophysically simplistic, albeit computationally sophisticated, surveys of what neutrino flavor and spin transformations might occur, by simulating neutrino spin and flavor evolution using a variety of potential supernova electron fraction profiles, absolute neutrino masses, and neutrino luminosities. In this paper, we present the results corresponding to one example set of parameters that led to large, measurable spin transformations. We look at the prospect for these spin transformations in both two- and three-flavor--- coupled with two spin states--- simulations carried out using a single angle neutrino bulb geometry (see Sec.~\ref{sec:Hamiltonian}) with the correct geometric dilution of neutrino fluxes. 

In Sec.~\ref{sec:Hamiltonian} of this paper, we discuss the Hamiltonian used in both the flavor and spin evolution of the neutrinos as well as the geometry of the neutrino bulb model. In Sec.~\ref{sec:results} we present the results of our simulations, we discuss them in Sec.~\ref{sec:discussion}, and we conclude in Sec.~\ref{sec:conclusion}.

\section{Hamiltonian}\label{sec:Hamiltonian}

\begin{figure}[ht]
\centering
\includegraphics[width=.5\textwidth]{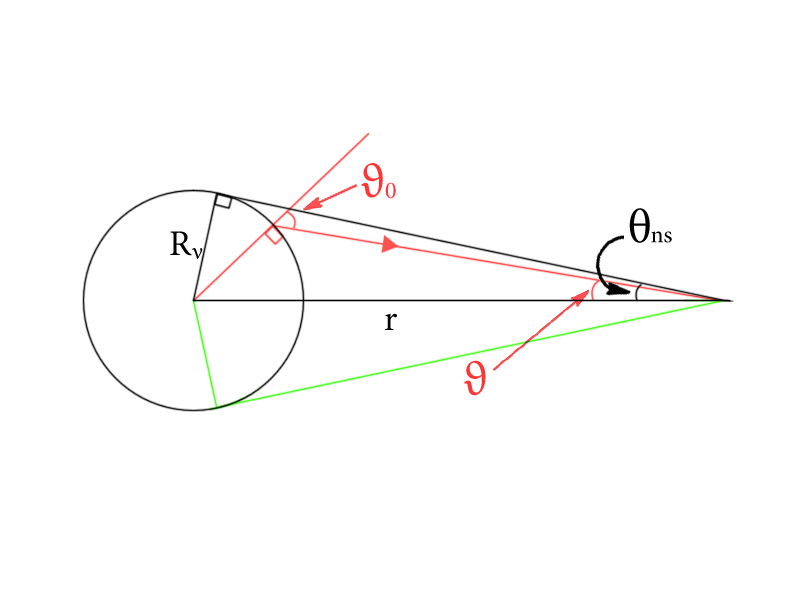}
\caption{Figure showing the basic geometry of the bulb model that we employ. Neutrinos are emitted isotropically from the surface of a central neutrino sphere with radius $R_\nu$, and subsequently interact with the matter background in the envelope and other neutrinos coming from this neutrino sphere.}
\label{fig:bulb}
\end{figure}

In this paper we consider the coherent evolution of neutrinos undergoing forward scattering on a matter background and a background of other neutrinos in a neutrino bulb model \cite{Duan06c,Duan07a,Duan:2010fr}. Electron neutrinos are assumed to be emitted isotropically from the surface of a central neutrino sphere (or \lq\lq bulb\rq\rq) of radius $R_\nu\approx 60\,\text{km}$ (see Fig.~\ref{fig:bulb}), with a Fermi-Dirac blackbody-shaped distribution of energies
\begin{equation} 
f(E_\nu)=\frac{1}{F_2(\eta_\nu)T_\nu^3}\frac{E_\nu^2}{e^{E_\nu/T_\nu-\eta_\nu}+1},
\end{equation}
where $\eta_\nu$ is the degeneracy parameter, and 
\begin{equation}
F_k(\eta_\nu)=\int_0^\infty\frac{z^k}{e^{z-\eta_\nu}+1}dz,
\end{equation}
so that the distribution is normalized,
\begin{equation}
\int_0^\infty f(E_\nu)dE_\nu = 1\quad .
\end{equation}

We first consider a two-flavor neutrino example. These considerations are generalizable to the three-flavor case in obvious fashion. Since we are considering coherent flavor and spin evolution, the neutrinos can be described as pure states in a four-component ket, with radius and neutrino energy, i.e., $(r,E_\nu)$, arguments suppressed for brevity \cite{1989neas.book.....B,2014AIPC.1594..313B,2016AIPC.1743d0001B}:
\begin{equation}
\begin{split}
\left|\Psi_{\nu_e}\right>&=\begin{pmatrix}1\\0\\0\\0\\
\end{pmatrix},\quad
\left|\Psi_{\nu_x}\right>=\begin{pmatrix}0\\1\\0\\0\\
\end{pmatrix},\quad \\
\left|\Psi_{\bar{\nu}_e}\right>&=\begin{pmatrix}0\\0\\1\\0\\
\end{pmatrix},\quad
\left|\Psi_{\bar{\nu}_x}\right>=\begin{pmatrix}0\\0\\0\\1\\
\end{pmatrix}.
\end{split}
\end{equation}

In this paper, when dealing with two-flavor situations, we will use the symbol \lq\lq $x$\rq\rq, in place of \lq\lq $\mu$\rq\rq\ or \lq\lq $\tau$\rq\rq\ flavors, to refer to the second flavor state (besides the electron flavor). The $\nu_x$ refers to a particular linear combination of the nearly maximally mixed $\nu_\mu$- and $\nu_\tau$-flavor states~\cite{Balantekin:2000hl,Caldwell:2000db}. The neutrinos obey a Schr\"odinger-like equation, which for a neutrino of energy $E_\nu$ is \cite{1989neas.book.....B, 2014AIPC.1594..313B,2016AIPC.1743d0001B,Cherry:2012lr, Duan06b,Duan06a,Duan07b}:
\begin{equation}
i\hbar\frac{\partial}{\partial r}|\Psi_{\nu}\rangle=H(r,E_\nu)|\Psi_{\nu}\rangle,
\end{equation}
where the Hamiltonian is now a $4\times 4$ matrix which encodes all the flavor and spin evolution of the neutrino states. In future discussion, we will also suppress the $(r,E_\nu)$ arguments in the Hamiltonian for brevity. For convenience of discussion, we break up the Hamiltonian into $2\times 2$ blocks:
\begin{equation}
H=\begin{bmatrix}
	H_{vac}+H_m+H_{\nu\nu} & H^{sf}\\
	(H^{sf})^\dagger & (H_{vac}-H_m-H_{\nu\nu})^T
\end{bmatrix}.
\label{eqn:fullHamiltonian}
\end{equation}

\subsection{Diagonal Hamiltonian}

The diagonal blocks of the total Hamiltonian are familiar from normal flavor evolution physics, with the caveat that the diagonal entries of $H_m$ and $H_{\nu\nu}$ now have to be defined relative to the vacuum rather than relative to other flavors. Another way to state this is to say that the traces which were removed from $H_m$ and $H_{\nu\nu}$ in usual studies of flavor evolution now have to be restored.

First we look at the vacuum term $H_{vac}$ which is the Hamiltonian arising merely from the fact that neutrino mass eigenstates are not coincident with neutrino flavor eigenstates \cite{1989neas.book.....B}. The vacuum Hamiltonian for both the neutrino sector and the antineutrino sector are the same since neutrinos and antineutrinos have the same mass \cite{1989neas.book.....B,Duan06a}:
\begin{equation}
H_{vac}=\frac{\delta m^2}{4E_\nu}U\begin{bmatrix}-1 & 0 \\ 0 & 1\end{bmatrix}U^\dagger .
\label{eqn:hvac}
\end{equation}

Note that here we can still use the traceless version of the vacuum Hamiltonian in this case. Here $\delta m^2=m_{\nu,2}^2-m_{\nu,1}^2$ is the mass-squared splitting of the two neutrino species, which we have taken to be either the solar splitting $\delta m^2=\delta m_\odot^2=7.6\times 10^{-5}\,\text{eV}^2$ or the atmospheric splitting $\delta m^2=\delta m_{atm}^2=2.4\times 10^{-3}\,\text{eV}^2$ \cite{1989neas.book.....B}. $U$ is the two-flavor version of the Pontecorvo-Maki-Nakagawa-Sakata (PMNS) matrix \cite{1989neas.book.....B}:
\begin{equation}
U=\begin{bmatrix}\cos\theta_\text{V} & \sin\theta_\text{V} \\ -\sin\theta_\text{V} & \cos\theta_\text{V}\end{bmatrix}\begin{bmatrix}1 & 0 \\ 0 & e^{i\alpha/2}\end{bmatrix}.
\end{equation}

In this matrix, $\alpha$ is the Majorana phase which we have set to $\alpha=0$ (we find that the results are insensitive to $\alpha$, which is discussed in Sec.~\ref{sec:majorana}), and $\theta_\text{V}$ is the mixing angle which we have taken to be $\theta_\text{V}=8.7^\circ$ for two-flavor simulations. The three-flavor version of the PMNS matrix will have three mixing angles, a $CP$ violating Dirac phase, and two $CP$ violating Majorana phases. Note that, even if $\alpha\neq 0$ here, the matrix multiplication in equation \ref{eqn:hvac} will result in the cancellation of the Majorana phase terms. Equation \ref{eqn:hvac} will be unchanged by a change in the Majorana phase, and so, as expected, \textit{flavor} transformations are not affected by a Majorana phase. 

The diagonal block matter term $H_m$ is the same term familiar from neutrino flavor transformation physics, except that, as mentioned, the Hamiltonian must now be defined with respect to the vacuum. Therefore we must also include contributions from the neutral current scattering of neutrinos as well as charged current scattering \cite{2015PhLB..747...27C,1989neas.book.....B,Duan06a}:
\begin{equation}
H_m=\sqrt{2}G_F(1-V_{out}\cos\beta)\begin{bmatrix}n_e-n_n/2 & 0 \\ 0 & -n_n/2\end{bmatrix},
\end{equation} 
where $G_F$ is the Fermi weak coupling constant, $n_e$ is the local net electron number density, $n_e \equiv n_{e^-}-n_{e^+}$, and $n_n$ is the local neutron number density. $V_{out}$ is the local outflow velocity of matter and $\beta$ is the angle the neutrino makes with the matter outflow. Due to net charge neutrality, we can express this Hamiltonian in terms of the baryon number density $n_b$ and the electron fraction $Y_e\equiv n_e/n_b$ \cite{2015PhLB..747...27C, 1989neas.book.....B,Duan06a}:
\begin{equation}
H_m=\frac{G_F n_b}{\sqrt{2}}(1-V_{out}\cos\beta)\begin{bmatrix}3Y_e-1 & 0 \\ 0 & Y_e-1\end{bmatrix}.
\end{equation} 

The diagonal block neutrino-neutrino Hamiltonian $H_{\nu\nu}$ is more complicated and will depend on the geometry of the neutrino trajectories. Again, we have to define this Hamiltonian with respect to the vacuum. For a bulb model, the neutrino-neutrino Hamiltonian is \cite{2015PhLB..747...27C,Duan06a}:
\begin{widetext}
\begin{equation}
H_{\nu\nu}=\frac{\sqrt{2}G_F}{2\pi R_\nu^2}\sum_{\kappa}\int_0^\infty\int_0^{\theta_{ns}}\frac{L_{\nu,\kappa}}{\langle E_{\nu,\kappa}\rangle}(1-\cos\vartheta\cos\vartheta')\Lambda_{\nu,\kappa}(E',\vartheta') f_{\nu,\kappa}(E')\sin\vartheta'd\vartheta'dE' .
\label{eqn:hvv}
\end{equation}
\end{widetext}

Here, the index $\kappa$ refers to the flavor and spin state of neutrinos \textit{at the point of emission}, i.e., at the neutrino sphere surface. $\kappa$ runs over all four of the flavor and spin states; i.e., $\kappa=1$ is an electron neutrino, $\kappa=2$ is an $x$-neutrino, $\kappa=3$ is an electron antineutrino, and $\kappa=4$ is an $x$-antineutrino. The assumption that neutrinos are emitted in flavor and spin eigenstates is predicated on neutrino decoupling being instantaneous at the neutrino sphere, which is a reasonable approximation given the steep density profile. $\theta_{ns}$ is the maximum angle that the neutrino sphere subtends at the location of the neutrino which we are tracking, and thus $\sin\theta_{ns}=R_\nu/r$. $L_{\nu,\kappa}$ is the luminosity of the $\kappa$ state neutrinos emitted at the neutrino sphere, and $\langle E_{\nu,\kappa}\rangle$ is the average energy of those neutrinos. The angle $\vartheta$ is the angle that the test neutrino makes with the radial direction at the interaction site and we have to integrate over all the other neutrinos. Finally, $\Lambda_{\nu,\kappa}(E',\vartheta')$ is a two-by-two matrix:
\begin{widetext}
\begin{equation}
\Lambda_{\nu,\kappa}(E',\vartheta')=\begin{bmatrix}
2\rho_{ee,\kappa}+\rho_{xx,\kappa} & \rho_{ex,\kappa} \\ \rho_{ex,\kappa}^\star & \rho_{ee,\kappa}+2\rho_{xx,\kappa}
\end{bmatrix}(E',\vartheta') - \begin{bmatrix}
2\rho_{\bar{e}\bar{e},\kappa}+\rho_{\bar{x}\bar{x},\kappa} & \rho_{\bar{e}\bar{x},\kappa} \\ \rho_{\bar{e}\bar{x},\kappa}^\star & \rho_{\bar{e}\bar{e},\kappa}+2\rho_{\bar{x}\bar{x},\kappa} 
\end{bmatrix}(E',\vartheta') .
\label{eqn:lambdamatrix}
\end{equation}
\end{widetext}

The density matrix elements in this equation are defined from the pure state kets as follows:
\begin{equation}
\rho_{ij,\kappa}(r)= \Psi_{\nu,\kappa i}^\star(r)\Psi_{\nu,\kappa j}(r) \, .
\end{equation}  
Here, $\Psi_{\nu,\kappa i}$ is the $i$th component of the state ket of the neutrino which started out at the neutrino sphere in the $\kappa$ state. Here, the index $i$ runs over the same flavor/spin basis states as the index $\kappa$. Finally, since we are performing single angle calculations in this paper, the angle integrals can be evaluated analytically (for a spherical geometry). Since the single angle approximation entails that all neutrinos on all trajectories are assumed to evolve in the same way as a neutrino on the test trajectory, the density matrices are assumed to be not angle dependent, i.e., $\Lambda_{\nu,\kappa}(E',\vartheta')=\Lambda_{\nu,\kappa}(E')$. Therefore, we find \cite{Duan06a}
\begin{equation}
\begin{split}
H_{\nu\nu}=\frac{\sqrt{2}G_F}{2\pi R_\nu^2}\sum_{\kappa}\int_0^\infty\frac{L_{\nu,\kappa}}{\langle E_{\nu,\kappa}\rangle}(A(r)-B(r)\cos\vartheta)\\
\times\Lambda_{\nu,\kappa}(E') f_{\nu,\kappa}(E')dE'.
\end{split}
\end{equation}
Here we have defined
\begin{equation}
A(r)=1-\sqrt{1-\frac{R_\nu^2}{r^2}},\quad B(r)=\frac{1}{2}\frac{R_\nu^2}{r^2}.
\end{equation}

\subsection{Off-diagonal Hamiltonian}

In this subsection we discuss the off-diagonal block, the spin-flip Hamiltonian $H^{sf}$. This Hamiltonian consists of two parts, one due to a matter background $H_m^{sf}$ and one due to the other background neutrinos $H^{sf}_{\nu\nu}$. The total spin-flip Hamiltonian is \cite{Vlasenko:2014lr,2015IJMPE..2441009V,2015PhLB..747...27C, 2015PhRvD..91l5020K}:
\begin{equation}
H^{sf}=(H^{sf}_m+H^{sf}_{\nu\nu})\frac{m^\star}{E_\nu}+\frac{m^\star}{E_\nu}(H^{sf}_m+H^{sf}_{\nu\nu})^T
\end{equation} 
The spin-flip Hamiltonian, unlike the diagonal flavor evolution parts of the total Hamiltonian, depends on the absolute mass of the neutrino. The mass matrix $m$ is
\begin{equation}
m=U^\star\begin{bmatrix}
	m_1 & 0 \\
	0 & m_2 \\
\end{bmatrix}U^\dagger
\label{eqn:massmtx}
\end{equation}
and due to the presence of the $U^\star$ instead of $U$ in this transformation, the Majorana phase can have an effect on this mass term, and therefore on spin transformations. As one can clearly see, the $m^\star/E$ term will tend to make the spin-flip Hamiltonian much smaller than the diagonal block matter and neutrino-neutrino Hamiltonians. The matter and neutrino-neutrino parts of the spin-flip Hamiltonian are \cite{2015PhLB..747...27C, 2015PhRvD..91l5020K,Vlasenko:2014lr,2015IJMPE..2441009V}
\begin{equation} \label{eq:Hsfm}
H^{sf}_m=-\frac{G_F n_b}{2\sqrt{2}}V_{out}\sin\beta\begin{bmatrix}3Y_e-1 & 0 \\ 0 & Y_e-1\end{bmatrix}
\end{equation}
\begin{equation} \label{eq:Hsfnu}
\begin{split}
H^{sf}_{\nu\nu}=-\frac{\sqrt{2}G_F}{2\pi R_\nu^2}\sum_{\kappa}\int_0^\infty\int_0^{\theta_{ns}}\frac{L_{\nu,\kappa}}{\langle E_{\nu,\kappa}\rangle}\sin\vartheta\cos\vartheta'\\
\times\Lambda_{\nu,\kappa}(E',\vartheta') f_{\nu,\kappa}(E')\sin\vartheta'd\vartheta'dE'
\end{split}
\end{equation}
and again, in the single angle approximation we can perform the $\vartheta'$ integral in the last equation to obtain \cite{2015PhLB..747...27C}:
\begin{equation} \label{eq:Hsfnusa}
\begin{split}
H^{sf}_{\nu\nu}=\frac{\sqrt{2}G_F}{2\pi R_\nu^2}\sum_{\kappa}\int_0^\infty\frac{L_{\nu,\kappa}}{\langle E_{\nu,\kappa}\rangle}B(r)\sin\vartheta \\
\times\Lambda_{\nu,\kappa}(E') f_{\nu,\kappa}(E')dE'.
\end{split}
\end{equation}  

A nonzero spin-flip potential means that propagating neutrinos are, in general, coherent superpositions of left-handed and right-handed states. In other words, a neutrino's instantaneous energy eigenstates are not coincident with the neutrino's helicity eigenstates. As a neutrino propagates through the supernova environment its spin can rotate, for example, from an initial left-handed neutrino into a right-handed antineutrino.

\section{Results}\label{sec:results}

In this study, we ran several single angle simulations with a variety of initial conditions and neutrino parameters. An example set of conditions and parameters which fostered significant spin-flip transformations are outlined in Table \ref{table:params}, and the corresponding results are presented below. We used a version of the flavor evolution code developed by the authors in Refs. \cite{Duan06a,Duan06b,Duan06c,Duan07a,Duan07b,Duan07c,Duan08,Duan:2008qy,Duan:2008eb,Cherry:2010lr,Duan:2010fr}, but extensively modified to incorporate the spin degrees of freedom described above. We used conditions which are similar to those found during the neutronization burst epoch of a supernova: very high electron neutrino luminosities and no other flavor or spin states (i.e. no antineutrinos) present. The luminosity we used is toward the higher end of possible luminosities even for the neutronization burst, but this high luminosity was necessary to obtain significant spin transformations. As the neutronization burst neutrinos were emitted from the core before the shock front had traversed the material in the envelope of the star, matter speeds were subsonic and therefore we took the outflow velocity to be zero. 

\begin{table}[h]
\begin{ruledtabular}
\begin{tabular}{lcdr}
\text{Parameter} & 
\text{Value} \\
\colrule
	$L_{\nu,e}$ & $1.8\times 10^{54}\, \text{erg/s}$\\
   	$L_{\nu,x,\bar{e},\bar{x}}$ & $0\,\text{erg/s}$\\
	$\langle E_{\nu,e}\rangle$ & $11\, \text{MeV}$\\
	$V_{out}$ & $0\, \text{m/s}$\\
	$m_1$ & $10\, \text{eV}$\\
	$\delta m^2_\odot$ & $7.6\times 10^{-5}\,\text{eV}^2$\\
	$\delta m^2_{atm}$ & $2.4\times 10^{-3}\, \text{eV}^2$\\
	$\vartheta_0$ & $60^\circ$\\
	$\theta_{12}$ & $34.4^\circ$\\
	$\theta_{13}$ & $8.7^\circ$\\
	$\theta_{23}$ & $45^\circ$\\
	$\delta_{cp}$ & $0$\\
	$\alpha$ & $0$\\

\end{tabular}
\caption{Parameters used in single angle simulations with spin-flip. The parameters are chosen to highlight the spin-flip effect and also to match, as much as possible, the neutronization burst epoch of a O-Ne-Mg supernova. For two-flavor simulations we used the mixing angle $\theta_{\text{V}}=\theta_{13}$.}
\label{table:params}
\end{ruledtabular}
\end{table}

The spin-flip potential experienced by a test neutrino is proportional to the component of the matter and neutrino currents transverse to its momentum. In a spherically symmetric model, for a radially directed test neutrino, the transverse background neutrino current has to add up to zero, just by symmetry (as is reflected in the $\sin \vartheta$ dependence in equation~\ref{eq:Hsfnu}). Therefore, in the absence of convective currents or asymmetric matter outflows, a radially directed neutrino would experience no spin-flip potential. Consequently, we have chosen to track a neutrino which is emitted at 60 deg ($\vartheta_0 = 60^\circ$) with respect to the normal (radial direction) of the neutrino sphere. In a realistic supernova model, the presence of transverse currents and asymmetries in the neutrino outflow could, in principle, give rise to a spin-flip effect in even radially directed neutrinos.    

\begin{figure}[hbt]
\centering
\includegraphics[width=.5\textwidth]{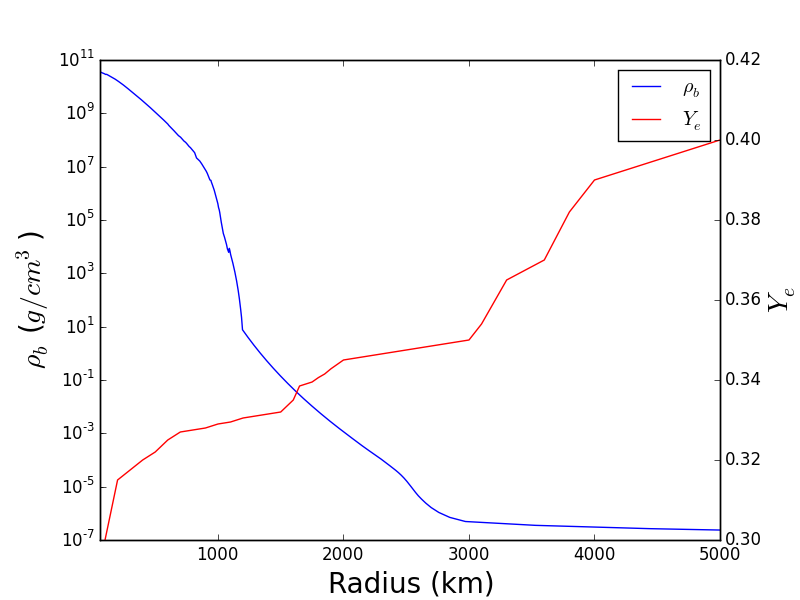}
\caption{The baryon density (blue) and electron fraction (red) profiles that we used in our simulations. The density profile is that of an O-Ne-Mg supernova taken from \cite{Cherry:2013lr,2010A&A...517A..80F}, while the electron fraction profile was created artificially so as to increase the chances of inducing significant spin transformations. Notice that the density profile is extremely centrally concentrated, with a steep dropoff at $r \approx 1000$ km.}
\label{fig:densprof}
\end{figure}

Figure \ref{fig:densprof} shows the baryon density, $\rho_b$, and electron fraction, $Y_e$, profile we used for our simulations. The electron fraction was set to hover close to $Y_e\approx 1/3$ relatively close in to the neutrinosphere so as to best facilitate the spin transformations (see section \ref{sec:spinresonance} for details on why this is). Electron fraction profiles which were not flattened near $Y_e\approx 1/3$ or which go through $Y_e\approx 1/3$ significantly farther out did not produce a significant spin transformation effect. It should be noted that, even though the electron fraction profile we used in this study is artificial, it does, however, conform to the general expectation that the electron fraction is lower closer to the neutron rich material in the core and grows as we move out into the envelope. 

\subsection{Solar Splitting}

Two-flavor simulations using the solar splitting, $\delta m^2_\odot$, were carried out first in order to get a feeling for the spin transformations. Two-flavor simulations using the solar splitting are significantly faster to run than ones which use the atmospheric splitting. With a larger mass-squared splitting like the atmospheric one, the natural flavor oscillation wavelength is much shorter and as such, the step sizes used in simulations become much smaller. Full three-flavor simulations are quite computationally intensive and take upward of eight hours or more to run. Solar splitting results are also much simpler in terms of the flavor evolution, and it is therefore easier to concentrate on the spin degrees of freedom. As such, most of the parameter space in terms of luminosities, electron fraction density profiles, etc., was explored using the solar splitting simulations. Only after finding significant spin transformations do we then run atmospheric splitting and three-flavor simulations in order to gauge any effect the spin transformations have on flavor transformations or vice versa. 

\begin{figure*}[ht]
\centering
\subfloat{\includegraphics[width=.5\textwidth]{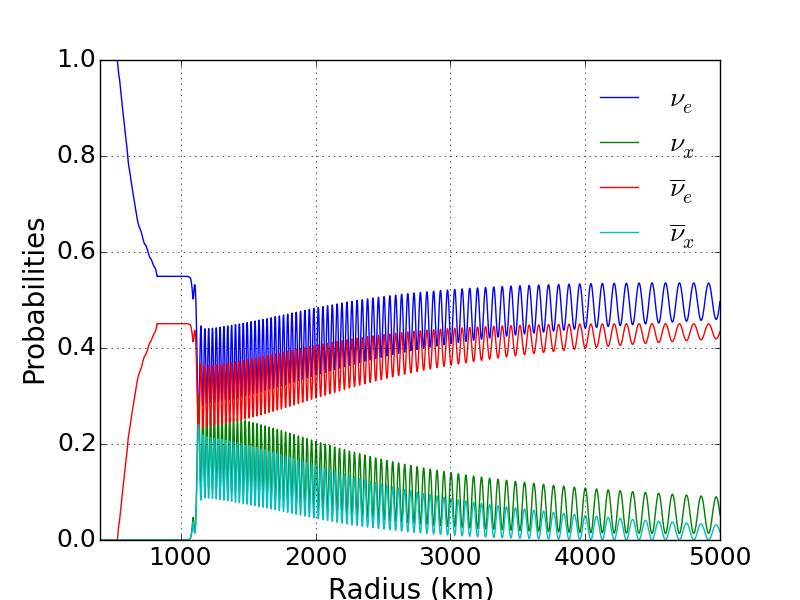}}
\subfloat{\includegraphics[width=.5\textwidth]{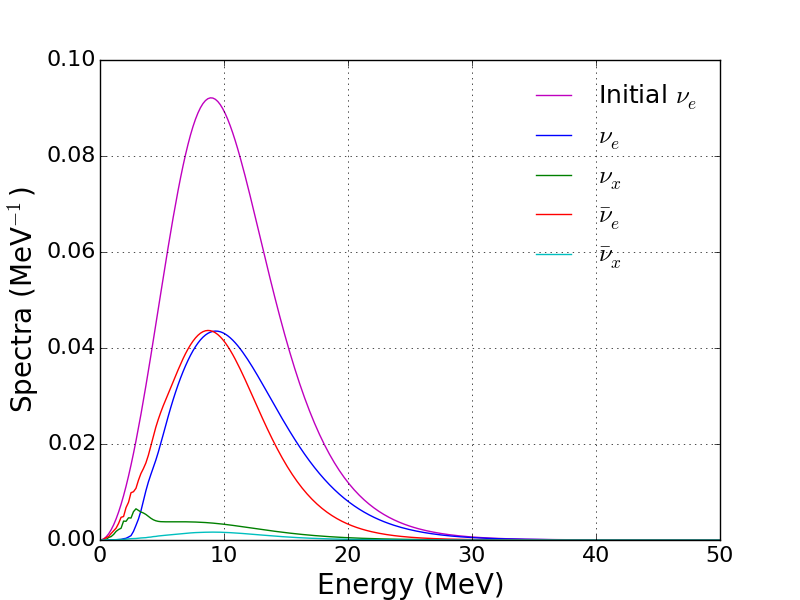}}
\caption{The left-hand graph shows the probabilities for a neutrino which started out in the electron neutrino state to be in any of the four possible states as a function of radius. As spin-flip resonance occurs, beginning around a radius of $\approx 500 \, \text{km}$, a large percentage of electron neutrinos are converted into electron antineutrinos. The neutrino flavor states stay stable for a few hundred kilometers before flavor evolution begins at a radius of $\approx 1100 \, \text{km}$. The right-hand graph shows the normalized final neutrino energy spectral distribution functions. The normalization we employed here is the same as the normalization employed in \cite{Duan06a}. The area under the magenta initial curve and, therefore the sum of the areas under the other four colored curves are equal to 1.}
\label{fig:solarresult}
\end{figure*}

\begin{figure*}[ht]
\centering
\subfloat{\includegraphics[width=.5\textwidth]{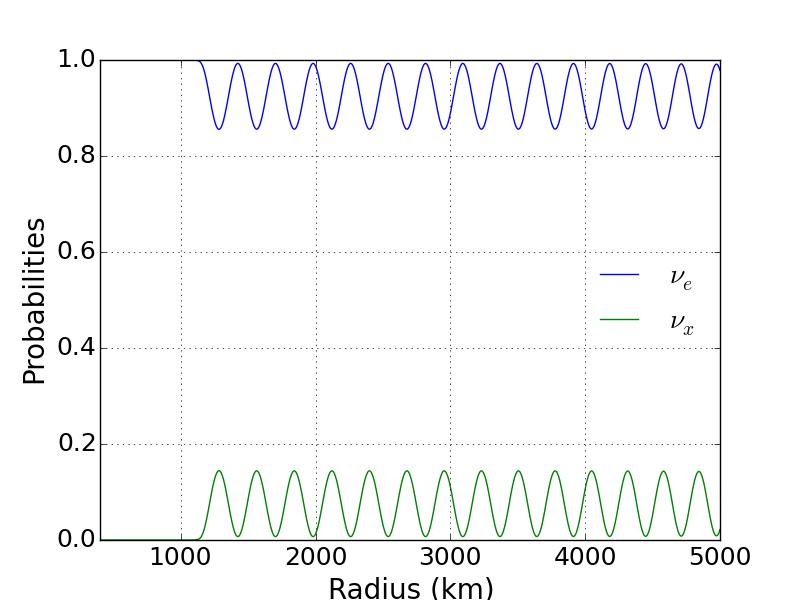}}
\subfloat{\includegraphics[width=.5\textwidth]{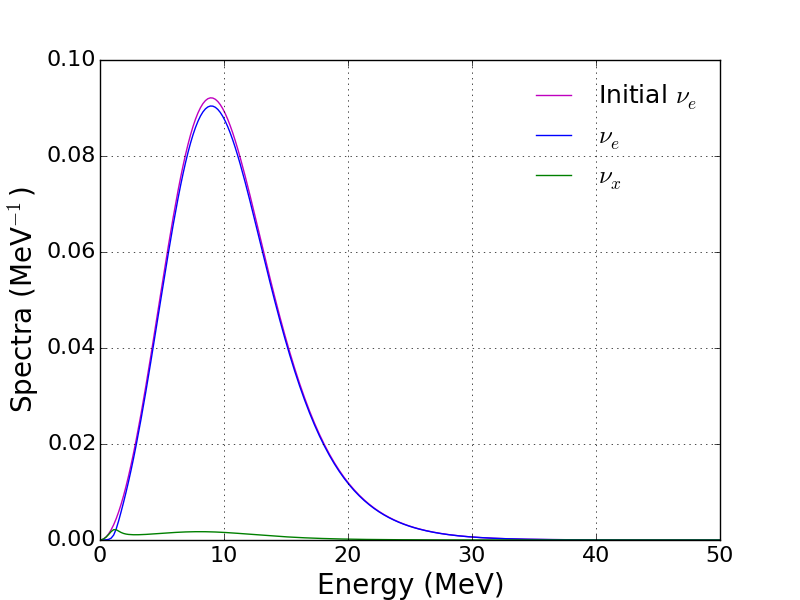}}
\caption{These are the probability evolution and spectral graphs for a solar splitting simulation where the spin-flip term has been turned off. All other parameters are the same as those used to produce the simulation in figure \ref{fig:solarresult}}
\label{fig:solarresultnsf}
\end{figure*}

Figure \ref{fig:solarresult} shows the energy averaged probability evolution history and the final spectral distribution  of the initial electron neutrinos propagating out from the supernova to a final simulation radius of $5000 \, \text{km}$. The final simulation distance of $5000 \,\text{km}$ was chosen to be quite far out so that we could see both the interesting flavor and spin transformations. We can see that significant spin transformation did occur. Approximately 45\% of initial electron neutrinos were converted into electron antineutrinos. Once the spin-flip transformation ends at a radius of about $r\approx 800 \, \text{km}$, the ratio of neutrinos to antineutrinos stays constant for a couple hundred kilometers, after which flavor transformations take over. The flavor evolution appears to go into a collective oscillation mode where essentially all of the neutrinos at all energies oscillate in step.

From the final spectral distribution we can see that the spin transformations converted preferentially lower energy neutrinos into antineutrinos, while leaving the very high energy neutrinos intact. This is to be expected simply due to the $m/E_\nu$ factor in the spin-flip Hamiltonian which suppresses the spin-flip for high energy neutrinos. However, the fact that the spin transformation was not limited to simply the lower energy bins, but affected the mean energy neutrinos as well, is an interesting result. The spin transformation converted neutrinos into antineutrinos, and then the flavor transformation gave rise to $x$ and $\bar{x}$ neutrinos. Consequently, although we started out with all electron neutrinos, by the end of our simulation, we had neutrinos of every flavor and spin.

Figure \ref{fig:solarresultnsf} shows the same graphs as Fig.~\ref{fig:solarresult}, for a simulation using the exact same parameters but with the spin coherence term turned off. The flavor transformations in Figs.~\ref{fig:solarresult} and \ref{fig:solarresultnsf} are qualitatively similar in the sense that, in both cases, beyond $r \approx 1000$ km, the neutrinos undergo synchronized flavor oscillations with a small amplitude, thereby largely preserving their flavor composition through the process. The frequency of synchronized oscillations, $\Omega_\text{sync}$, is higher in the presence of antineutrinos (shown in figure \ref{fig:solarresult}), as is expected~\cite{Duan:2010fr,Pastor:2002zl}. Note that synchronized oscillations with the solar neutrino mass-squared splitting are still in effect at our final radius of $r=5000 \, \text{km}$ due to the smallness of the solar neutrino mass-squared splitting and therefore the vacuum Hamiltonian. 

After running several simulations using the solar neutrino mass-squared splitting, we found that the spin transformations are very sensitive to the initial conditions inside the supernova. For example, our simulations have shown that, keeping everything else constant, raising or lowering the neutrino luminosity by more than 20\%-30\% from our adopted value will essentially destroy spin transformations. Additionally, if the electron fraction profile was made to go through $Y_e \lesssim 1/3$ more quickly, or if the neutrino rest mass was set to significantly less than the unrealistically large \cite{Abazajian:2011lr,2016arXiv161002743A, 2011PhRvD..84k2003A, 2015PhRvL.114p2501A,2005EPJC...40..447K,2016JPhCS.718b2013M} $10 \, \text{eV}$ value, no significant spin transformations occurred. 

This behavior can be explained as follows: in order to achieve significant spin transformation, nonlinear effects must take hold to keep the neutrino Hamiltonian near resonance (the so-called tracking behavior; see section \ref{sec:nonlinear} for details). Therefore, it is possible that a small to moderate change in initial conditions will drastically affect the spin transformations. If the neutrino Hamiltonian is not kept near resonance, and no tracking behavior develops, the conversion of neutrinos into antineutrinos becomes entirely negligible in terms of a potential terrestrial detection (usually of order one part in 10 billion).

If these spin transformations happened in a real supernova, the effects could be detectable. Without spin transformations we expect not to see a significant antineutrino content coming from a neutronization burst signal. Thus, just from the solar splitting simulations, we would now expect a significant antineutrino content which would be robust to any flavor transformation physics.

\subsection{Atmospheric Splitting}

\begin{figure*}[htb]
\centering
\subfloat{\includegraphics[width=.5\textwidth]{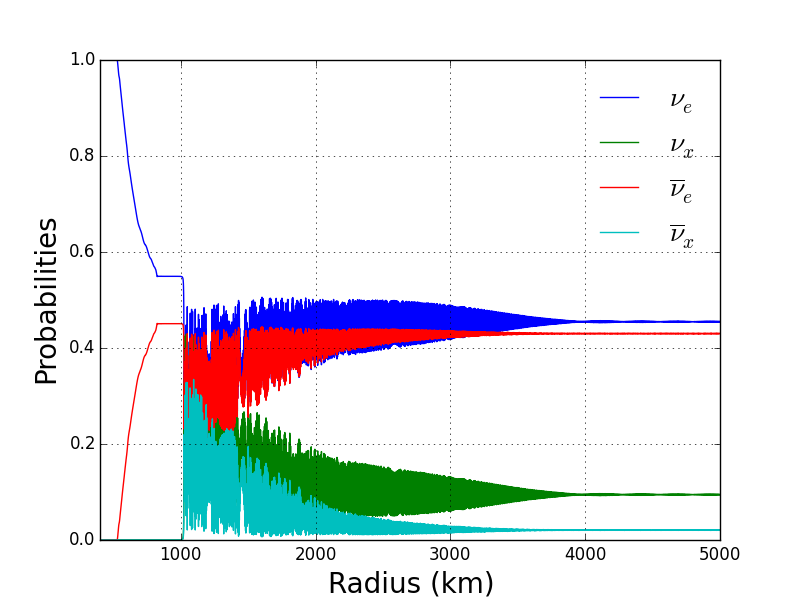}}
\subfloat{\includegraphics[width=.5\textwidth]{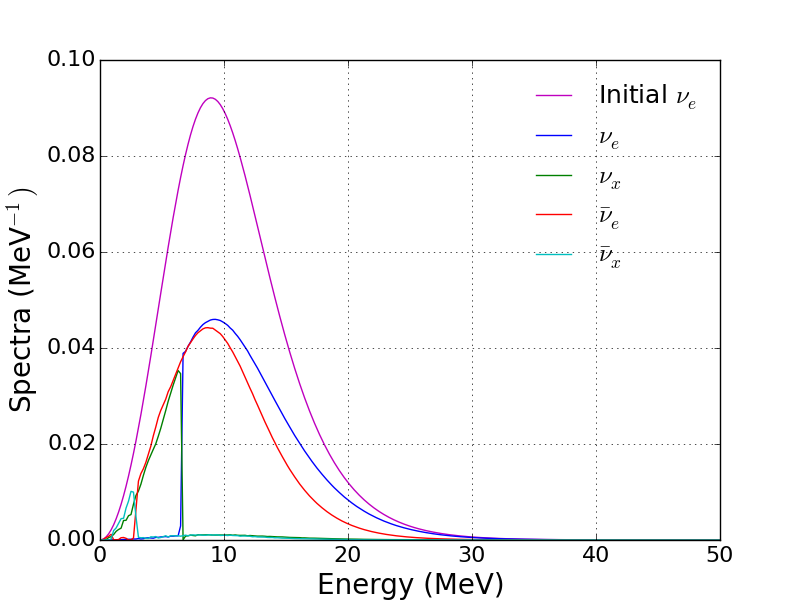}}
\caption{As with figure \ref{fig:solarresult}, the left-hand graph shows the evolution of a neutrino which started out in the electron neutrino state and the right-hand graph shows the normalized final neutrino energy spectral distribution functions. Here we see that the spin transformations, $\nu_e\rightarrow\bar{\nu}_e$, were not changed from the simulations with the solar mass-squared splittings.}
\label{fig:atmresult}
\end{figure*}

\begin{figure*}[htb]
\centering
\subfloat{\includegraphics[width=.5\textwidth]{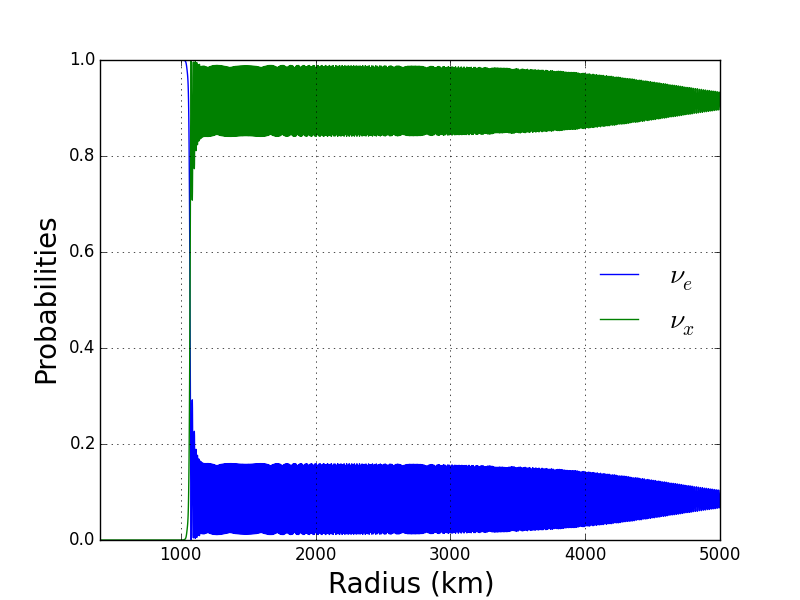}}
\subfloat{\includegraphics[width=.5\textwidth]{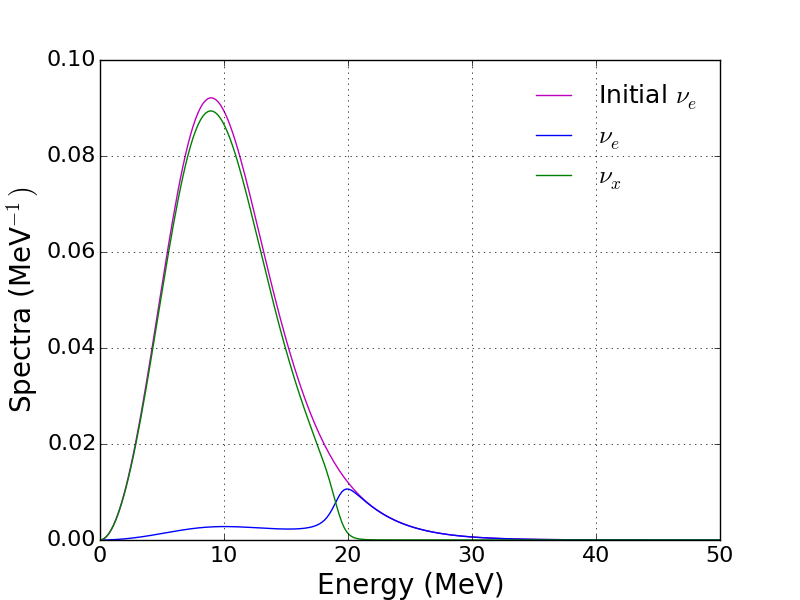}}
\caption{These are the probability evolution and spectral graphs for an atmospheric splitting simulation where the spin-flip term has been turned off preventing any possibility for the spin-flip. All other parameters are the same as those used to produce the simulation in figure \ref{fig:atmresult}.}
\label{fig:nospinatm}
\end{figure*}

Simulations performed with the atmospheric neutrino mass-squared splitting and full three-flavor simulations showed essentially the same spin-transformation phenomena as the simulations with the solar splitting given in the previous section. In a broad brush, only the flavor transformations differ among the different simulations. This makes sense because the absolute neutrino mass that we chose to analyze is several orders of magnitude larger than the mass splittings.  

Figure \ref{fig:atmresult} shows the results obtained from two-flavor simulations using the atmospheric mass-squared splittings instead of the solar ones. As we can see, the spin transformations proceeded essentially identically to the solar splitting results. Again, approximately 45\% of neutrinos were converted into antineutrinos and the spectrum of transformed neutrinos is the same as before. The spin-flip preferentially transformed lower energy neutrinos into antineutrinos. It is not surprising that in these simulations the spin transformations were not affected. The spin transformations occurred prior to any flavor transformations (for an examination of why, see section \ref{sec:onsetoftransformations}). The spin transformations began at a radius of $r\approx 500 \, \text{km}$, and all spin conversion was finished at a radius of $r\approx 800$ to the 55\%--45\% ratio of neutrinos to antineutrinos we see in the final spectrum. The flavor transformations did not set in until a radius outside of $r\gtrsim 1000 \, \text{km}$; this is in agreement with previous studies of flavor transformation in the neutronization epoch of an oxygen-neon-magnesium supernova\cite{Cherry:2013lr,Cherry:2010lr,Duan08}. Therefore, for these simulations, flavor transformations do not have a chance to feed back on the spin transformations.

The converse statement, however, is not true. Spin transformations in our simulations can have an effect on flavor transformations since they happen first. A transformation of 45\% of neutrinos into antineutrinos affects the diagonal blocks of the Hamiltonian significantly and can change the subsequent flavor evolution. Figure \ref{fig:nospinatm} shows the results of a simulation where the spin coherence term has switched off. No spin flip was allowed to occur and only flavor transformations were possible. Unlike the solar mass-squared splitting case, the flavor evolution in the atmospheric mass-squared splitting simulations were qualitatively affected by the spin transformations. These results presented in figure \ref{fig:nospinatm} differ qualitatively from those in figure \ref{fig:atmresult}. The flavor evolution here is qualitatively quite similar to previous studies of the flavor evolution for neutronization burst neutrinos in O-Ne-Mg supernovae\cite{Cherry:2010lr,Duan08}. Even though we have used a quite high neutrino luminosity, we still get significant flavor transformation from the electron neutrino state to the $x$-neutrino state, $\nu_e\rightarrow\nu_x$, for neutrinos with energies less than approximately $E_l\lesssim 20 \, \text{MeV}$, qualitatively similar to previous single angle and multiangle simulations. Almost all the low energy neutrinos have been converted by the so called \lq\lq neutrino-background-enhanced MSW-like flavor transformation\rq\rq\ \cite{Duan08}. By comparison, the results given in \ref{fig:atmresult} show a much lower threshold energy, $E_l\approx 9 \, \text{MeV}$, for the $\nu_e\rightarrow\nu_x$ flavor transformation channel. The presence of antineutrinos in the neutrino spectrum has affected the flavor transformations in such a dominant way. Without spin transformations, as much as approximately 90\% of neutrinos were transformed into the x-neutrino state, whereas with spin transformations, only about 20\% of the leftover neutrinos (those not transformed into antineutrinos) were converted into the $x$-neutrino state. 

For this atmospheric mass-squared splitting case, we made two movies to illustrate both the spin and the flavor transformations, which we show in the Supplemental Material~\cite{movies}. One can see in the spin coherence movie (titled \lq\lq Neutrino Spectra With Spin Coherence\rq\rq) explicitly the spin coherence developing around a radius of $r\approx 500 \, \text{km}$ starting with the lower energy neutrinos (see section \ref{sec:nonlinear} for a discussion of why this is the case). Synchronized flavor oscillations set in at a radius of $r\approx 1100 \, \text{km}$, and then a spectral swap develops beginning at a radius of $r\approx 2000 \, \text{km}$ (see section \ref{sec:flavorT} for a discussion of these flavor transformations). In the movie made with the spin coherence turned off (titled \lq\lq Neutrino Spectra Without Spin Coherence\rq\rq), one can see the neutrino background assisted MSW-like effect take hold at $r\approx 1100 \, \text{km}$, converting most electron neutrinos into $x$-neutrinos (again, see section \ref{sec:flavorT} for a discussion). Subsequently, synchronized flavor oscillations and then a spectral swap develop.  

\subsection{Three Flavor}

\begin{figure*}[ht]
\centering
\subfloat{\includegraphics[width=.5\textwidth]{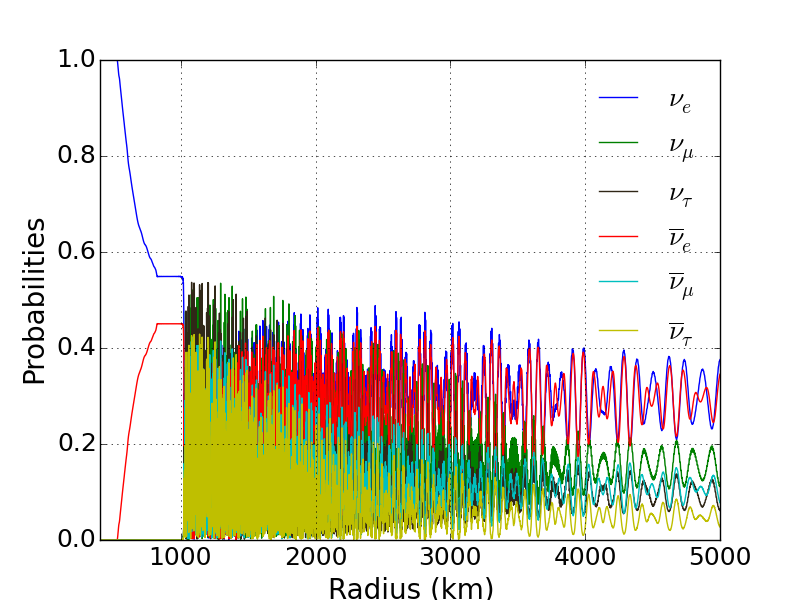}}
\subfloat{\includegraphics[width=.5\textwidth]{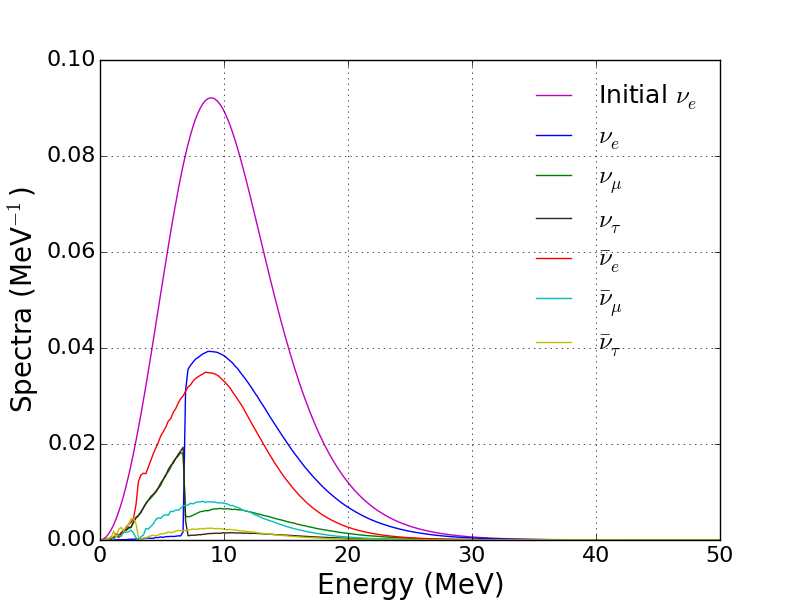}}
\caption{These are the probability evolution and spectral graphs for a full three-flavor simulation. As we can see, the flavor transformations are more complicated than two-flavor simulations but the spin transformations have not changed. Moreover, the flavor swaps $\nu_e\rightarrow\nu_\mu/\nu_\tau$ and $\bar{\nu}_e\rightarrow\bar{\nu}_\mu/\bar{\nu}_\tau$ at low energies are still present, making the three-flavor simulation results roughly a superposition of the two different two-flavor simulation results.}
\label{fig:3fresult}
\end{figure*}

\begin{figure*}[ht]
\centering
\subfloat{\includegraphics[width=.5\textwidth]{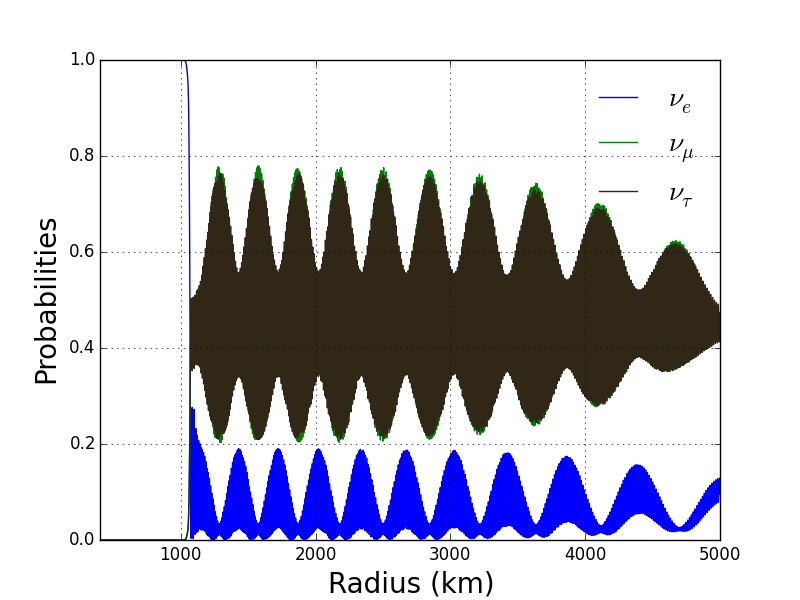}}
\subfloat{\includegraphics[width=.5\textwidth]{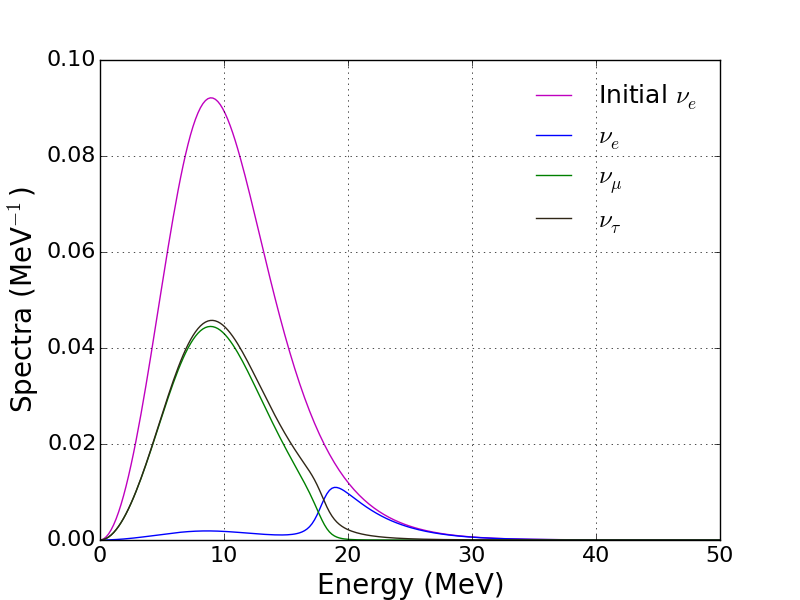}}
\caption{The same as figure \ref{fig:3fresult} but for a simulation without spin coherence. Again, the three-flavor results are roughly a superposition of the two separate two-flavor results where the $\mu,\tau$ flavors are maximally mixed and collectively act like the second $x$ flavor in the two-flavor simulations.}
\label{fig:3fnospinresult}
\end{figure*}

Finally, a full three-flavor simulation was run to see if any qualitative differences can be found in the spin coherence between a full three-flavor simulation and two-flavor simulations. For this three-flavor simulation, the $CP$ violating Dirac phase, $\delta_{CP}$, and both possible Majorana phases, $\alpha_1,\alpha_2$, were set to zero. The results are presented in figure \ref{fig:3fresult} for parameters which matched those used in figures \ref{fig:solarresult} and \ref{fig:atmresult}. Here we see again, not surprisingly, that the spin coherence has not been essentially changed at all from either of the two-flavor results. The flavor evolution appears to be a superposition of the two different mass-squared splitting results. We still have neutrinos of energy less than $E_l\approx 9 \, \text{MeV}$ being transformed into $\mu$ or $\tau$ neutrinos and we still have what appear to be collective neutrino oscillations like in the solar mass-squared splitting case although these oscillations are much messier here. The fact that three-flavor evolution is a superposition of the two two-flavor results is also consistent with previous three-flavor studies of the ONeMg neutronization burst \cite{Duan08}.

A three-flavor simulation with the spin coherence turned off was also run in order to compare the flavor transformation results. Results are presented in figure \ref{fig:3fnospinresult}. We see here that the flavor transformation is essentially still a superposition of the two different mass-squared splitting results. The qualitative difference in how the flavor transformation between a simulation with and without spin coherence arises is again a superposition of the differences we found in the two different mass-squared splitting cases. By turning on the spin coherence, the swap energy, $E_l$, moved from $\approx 20 \, \text{MeV}$ to $\approx 9 \, \text{MeV}$ just like in the atmospheric splitting case. The synchronized oscillation frequency, $\Omega_{sync}$, in the solar mass-squared splitting regime grew in the presence of antineutrinos just like in the two-flavor simulations using the solar mass-squared splitting.

Since the three-flavor results are a superposition of the two different two-flavor results, and since the $\mu$ and $\tau$ flavors are essentially maximally mixed, this lends credence to our separate two-flavor analyses with solar and atmospheric mass-squared splittings. For clarity and simplicity, then, we can choose to focus our discussions on the two-flavor simulations. The three-flavor simulations do not present any phenomenon that was not present in the two-flavor simulations.    

\section{Discussion} \label{sec:discussion}

\subsection{Spin Resonance Conditions}\label{sec:spinresonance}

In order for spin coherence to have a significant effect, the neutrinos must go through a resonance between left- and right-handed states~\cite{2014arXiv1406.6724V}. For our discussion here, we will restrict ourselves to the two-flavor case since three-flavor simulations did not differ in the spin evolution of the neutrinos from two-flavor simulations and since two-flavor neutrino evolution is much simpler and more intuitive. In flavor evolution, a MSW resonance occurs when the diagonal components of the Hamiltonian equal each other, i.e., when $H_{11}=H_{22}$ \cite{1989neas.book.....B}. For clarity, we note that some sources may simply state the resonance condition as $H_{11}=0$ for two-flavor (only) evolution due to the fact that removing the trace from a $2\times 2$ matrix means that $H_{11} = -H_{22}$ and so the two conditions are equivalent for a traceless $2\times 2$ Hamiltonian. Similarly, a resonance \cite{2015PhLB..747...27C, 2014arXiv1406.6724V} for the $\nu_e\rightleftharpoons\bar{\nu}_e$ channel happens when the $\nu\nu$-component (the 1-1 component) of the Hamiltonian is equal to the $\bar{\nu}\bar{\nu}$-component (the 3-3 component) of the Hamiltonian ($H_{11}=H_{33}$): 
\begin{equation}
(H_{vac}+H_m+H_{\nu\nu})_{11}=(H_{vac}-H_m-H_{\nu\nu})_{11}.
\end{equation}
This is taken directly from equation \ref{eqn:fullHamiltonian}. We can see as a consequence of the fact that neutrinos and antineutrinos have identical mass-squared splittings, i.e., they have the same rest mass, that the only way for this resonance condition to hold is if the 1-1 component of the matter Hamiltonian cancels out the 1-1 component of the neutrino-neutrino Hamiltonian:
\begin{equation}
\frac{G_F n_b}{\sqrt{2}}(3Y_e-1)+(H_{\nu\nu})_{11}=0 .
\end{equation} 
An immediate consequence of this resonance condition is that, unlike the classic MSW resonance~\cite{Wolfenstein78,Mikheyev85, 1989neas.book.....B}, it is not dependent on any neutrino energy. Neutrinos of all energies will go through this spin coherence resonance together. Close to the neutrino sphere, we expect the density to be so high that the matter term, neglecting the $(3Y_e-1)$ part, would dominate over the neutrino-neutrino term even with the extremely high neutrino luminosities seen during the neutronization burst. However, because the neutral current terms now contribute to an energy splitting between neutrinos and antineutrinos, into the Hamiltonian, we can see that $(H_m)_{11}$ will now be negative if $Y_e<1/3$ and positive if $Y_e>1/3$. Therefore, if $Y_e\approx 1/3$ the matter Hamiltonian can be suppressed relative to the neutrino-neutrino Hamiltonian. Indeed, we find that passing near $Y_e\lesssim 1/3$ is in fact necessary for resonance to occur close to the neutrino sphere. We need the electron fraction to be less than $1/3$ because the neutrino-neutrino Hamiltonian will be positive due to fact that there are not antineutrinos initially. The condition for resonance will be satisfied, then, as long as $Y_e(r=R_\nu)<1/3$ and then $Y_e$ passes through $1/3$ at some larger radius in our simulation. As discussed in Ref. \cite{2014arXiv1406.6724V}, the feedback physics in the spin resonance channel we discuss here is quite similar to the matter-neutrino resonance \cite{2012PhRvD..86h5015M,2016PhLB..752...89W, 2016PhRvD..93d5021M,2016PhRvD..93j5044V,2016arXiv160504903S, 2016arXiv160704671Z}.

\subsection{Adiabaticity}\label{sec:adiabaticity}
Although it seems quite likely that the Hamiltonian will pass through spin coherence resonance at some point, another extremely important aspect of the spin transformations, which is the same for flavor transformations in the MSW effect, is whether the system goes through resonance adiabatically or not. If $H_{11}$ goes through zero very quickly, very nonadiabatically, then one would expect no significant spin transformations will occur even though there is a resonance \cite{Duan07a,2016PhLB..752...89W,Qian95}. As a consequence, not only do we have to examine $H_{11}$, we of course also have to look at the spin-flip Hamiltonian itself. For the $\nu_e\rightleftharpoons\bar{\nu}_e$ channel, the relevant term to examine is $(H^{sf})_{11}$. Looking at this problem through the eyes of the MSW effect, we can define an adiabaticity parameter \cite{Duan07a,1989neas.book.....B,2016PhLB..752...89W,Qian95,2014arXiv1406.6724V}:
\begin{equation}
\gamma\equiv \left(\frac{2|(H^{sf})_{11}|^2}{\dot{H}_{11}}\right)_\text{res},
\label{eq:adiabat}
\end{equation}
where the subscript \lq\lq res\rq\rq\ indicates that the quantities on the right-hand side are being evaluated as the system is passing through resonance. The adiabaticity parameter must satisfy $\gamma\gg 1$ in order for the Hamiltonian to be considered adiabatic as far as spin transformations are concerned \cite{Qian95}. In other words, we want the spin-flip Hamiltonian to be large compared to the rate of change of the diagonal Hamiltonian term at resonance. We can immediately see, however, that due to the $m/E_\nu$ term in the spin-flip Hamiltonian, this adiabaticity condition will be hard to meet for the quickly changing conditions inside a O-Ne-Mg supernova. The extremely steep density dropoff and the geometric dilution of the neutrino fluxes make it especially hard for spin-flip transformations to be significant. Indeed, our simulations have so far been unsuccessful in generating large spin-flip transformations for neutrino masses $m_\nu \ll 10\,\text{eV}$. An iron core collapse supernova density profile would not be so centrally concentrated and might be better in terms of adiabaticity. Perhaps with an iron core collapse density profile, we could have gotten significant spin transformations for a smaller neutrino rest mass. For this paper, however, we chose to use the O-Ne-Mg supernova profile so that we could compare our flavor transformation results for the neutronization burst with previous studies like in \cite{Cherry:2010lr,Duan08}.

Of course, this neutrino mass of $\approx 10\, \text{eV}$ is unrealistically high. However, equation \ref{eq:adiabat} shows that the adiabaticity parameter would be increased by decreasing $\dot{H}_{11}$, the rate at which the diagonal Hamiltonian changes. A sufficiently flattened matter potential $\propto n_b(Y_e-1/3)$ could make spin transformations possible for more realistic neutrino masses, e.g. for $m_\nu \approx 0.1\, \text{eV}$. It must be noted here that although we did not artificially flatten the O-Ne-Mg supernova density profile for our simulations, we did use an electron fraction profile which hovered near $Y_e\lesssim 1/3$ for several hundred kilometers. In addition, as the neutrino-neutrino contribution to the Hamiltonian includes geometric dilution, that part of the Hamiltonian cannot be flattened.

As we move farther from the neutrino sphere, the $B(r)$ term in $H^{sf}_{\nu\nu}$, which encapsulates the integral over $\cos\vartheta'$, in the spin-flip Hamiltonian obviously drops by a factor of $r^2$. On top of that, the $\sin\vartheta$ term will drop as well for all emission angles $\vartheta_0$ as we move out from the neutrino sphere \cite{Duan07a,Duan06a}: 
\begin{equation}
\sin\vartheta=\frac{R_\nu\sin\vartheta_0}{r}
\end{equation}
As a consequence, geometric dilution means $H^{sf}_{\nu\nu}\propto1/r^3$. Notice that in equation \ref{eq:adiabat}, the adiabaticity parameter has an $|(H^{sf})_{11}|^2$ term in the numerator which will drop as six powers of the radius. Since this term drops so drastically as we get farther from the neutrinosphere, and it started out very small in the first place, it will be harder at large radius for the spin coherence to be adiabatic unless $\dot{H}_{11}$ is extremely flat. Even if $(\dot{H}_m)_{11}$ is extremely flat far from the neutrinosphere, $(\dot{H}_{\nu\nu})_{11}$ is determined simply by the geometric dilution of neutrinos. This term will certainly not decrease as six powers of the radius. Notice simulations have so far only been successful in obtaining significant spin transformations fairly close to the neutrino sphere as in the results presented in figure \ref{fig:solarresult}. A simple order of magnitude estimate from equation \ref{eq:adiabat} however, suggests that adiabaticity is unlikely to ever hold for spin transformations. Clearly nonlinear effects are needed in order to obtain a large spin transformation (see the following subsection). 

The $m/E_\nu$ in the spin-flip Hamiltonian has an additional effect in that it makes lower energy neutrinos go through resonance marginally more adiabatically than high energy neutrinos. Thus, if spin transformations do occur, we might expect that lower energy neutrinos are more preferentially transformed into antineutrinos than higher energy neutrinos. As mentioned earlier, all neutrinos will go through the spin resonance together, and it is only the adiabaticity of the resonance that changes between neutrinos of different energies. This fact could help explain why in figure \ref{fig:solarresult} the antineutrino spectrum arising from the spin coherence effect appears to be smooth and shows no sharp or jagged cutoffs in energies.


\subsection{Non-Linear Effects}\label{sec:nonlinear}

\begin{figure*}[hbt]
\centering
\subfloat{\includegraphics[width=.5\textwidth]{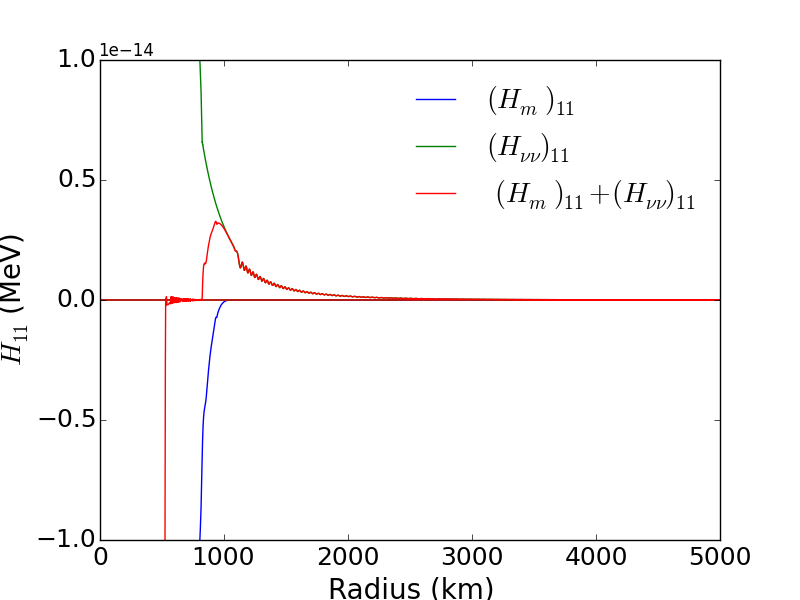}}
\subfloat{\includegraphics[width=.5\textwidth]{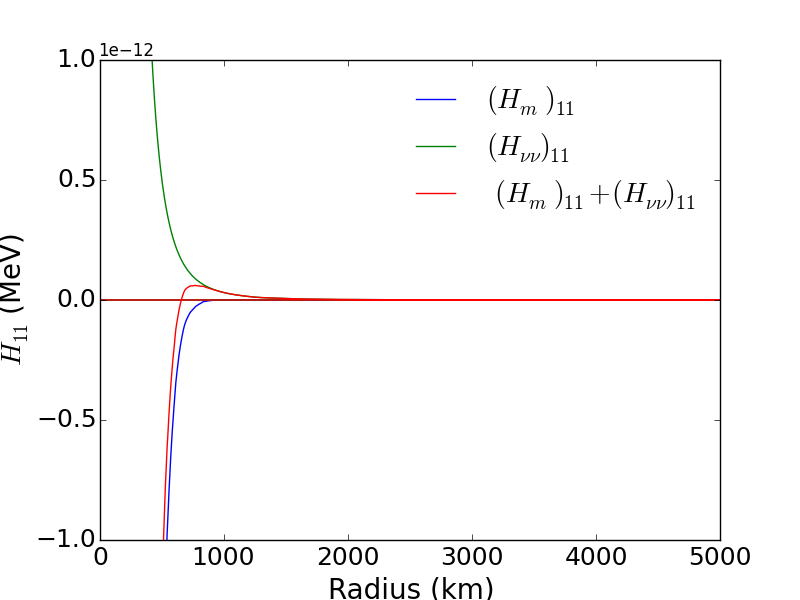}}
\caption{$(H_m)_{11}$, and $(H_{\nu\nu})_{11}$ as a function of radius. The green line is the neutrino-neutrino part of the Hamiltonian, the blue line is the matter part of the Hamiltonian, and the red line is the sum. Spin-flip resonance in the $\nu_e\rightleftharpoons\bar{\nu}_e$ channel occurs when the sum of these two elements of the Hamiltonian, the red line \lq\lq sum\rq\rq, is zero.  On the left-hand figure, this begins around a radius of $r\approx 500 \, \text{km}$ and ends around a radius of $r\approx 800 \, \text{km}$ which corresponds exactly to when the spin transformations in figure \ref{fig:solarresult} began and ended.  Due to nonlinear effects, as we can see, this element of the Hamiltonian tracks $H_{11}\approx 0\,\text{MeV}$ for several hundred kilometers. On the right-hand figure, which is for a simulation which did not produce significant spin transformation, the Hamiltonian does not appear to track $H_{11}\approx 0$. The right-hand simulation used all the same parameters as the left-hand simulation but with a steeper electron fraction profile.}
\label{fig:sfHamiltonian}
\end{figure*}

Nonlinear effects could strengthen spin transformations. Indeed, the simulations have shown that under specific circumstances, a large spin transformation effect can occur even if the transformation is not expected to be adiabatic for all but the most low energy neutrinos (as discussed above, our simulations require an unrealistically high neutrino rest mass and a specifically tailored electron fraction profile to obtain significant spin transformations). Naive linear reasoning, like that in section \ref{sec:adiabaticity} would lead us to the conclusion that even with the highly flattened electron fraction profile we used, no significant spin transformations should occur. However, as the neutrinos move through the resonance (for a growing electron fraction profile this will correspond to $(H_m)_{11}+(H_{\nu\nu})_{11}$ passing from negative to positive values), if some low energy neutrinos do transform into antineutrinos (as was the case for our spin coherence simulation \cite{movies}), this will tend to drive $(H_{\nu\nu})_{11}$ to lower values. If the rate of change of this effect is large enough, it can counteract the steeply rising matter potential, thus driving the sum of $(H_{\nu\nu})_{11}$ and $(H_m)_{11}$ back near zero. As discussed in \cite{2014arXiv1406.6724V}, this nonlinear feedback of the neutrino-neutrino interaction tends to keep the sum of $(H_{\nu\nu})_{11}$ and $(H_m)_{11}$ near zero. This nonlinear feedback forcing $(H_m)_{11}+(H_{\nu\nu})_{11}\approx 0$ for an extended length scale is what we mean by tracking behavior. 

The left-hand graph in figure \ref{fig:sfHamiltonian} shows $(H_m)_{11}$, $(H_{\nu\nu})_{11}$, as well as the sum $(H_m)_{11} + (H_{\nu\nu})_{11}$, for a neutrino as it evolves with radius in the simulation shown in figure \ref{fig:solarresult} (i.e., a simulation that produced a large spin-transformation effect). The right-hand graph in figure \ref{fig:sfHamiltonian} shows the same thing but for a simulation which showed no significant spin transformations. In that simulation, the electron fraction profile was not made to be extremely flat through resonance. All other parameters we kept the same. As such, an extreme lack of adiabaticity in the $\nu_e\rightleftharpoons \bar{\nu}_e$ transformation channel precluded even the nonlinear tracking from beginning and no significant spin transformations occurred. We can see that the diagonal Hamiltonian has been made to track $(H_m)_{11}+(H_{\nu\nu})_{11}\approx 0$ for a few hundred kilometers for the simulation which produced significant spin transformations, but for the simulation which did not produce significant spin transformations, $(H_m)_{11}+(H_{\nu\nu})_{11}$ simply passed through zero smoothly. We note that since $(H_m)_{11}+(H_{\nu\nu})_{11}$ did pass through zero even for the simulation shown in the right-hand graph in figure \ref{fig:sfHamiltonian}, the neutrinos did go through the spin resonance. The neutrinos simply did not go through the resonance adiabatically enough even for low energy neutrinos, and no tracking behavior was initiated, and thus no significant spin transformations occurred. We note that, although the electron fraction was set to hover near $Y_e\lesssim 1/3$ for the simulation shown in figure \ref{fig:solarresult}, that alone is not enough to force the Hamiltonian to track near zero for so long as we see in the left-hand graph in figure \ref{fig:sfHamiltonian}. The electron fraction was flattened but was not finely tuned in order to exactly cancel out the neutrino-neutrino potential. The tracking must be introduced by nonlinear effects in the neutrino evolution. We found over all simulations that this tracking behavior we just described is necessary in order to produce significant spin transformations. The tracking behavior was difficult to attain for various choices of parameter values. It seems likely that additional sources for feedback phenomena, e.g. electron fraction feed back mechanisms, might be necessary to increase the likelihood of getting into the tracking regime, and thereby produce a large spin-flip effect more robustly. 

\subsection{Majorana Phase}\label{sec:majorana}

Spin transformation calculations like ours with two or more neutrino flavors involve Majorana phases in a nontrivial way. For a two-flavor system of neutrinos, there can be one Majorana phase which can affect the spin-flip Hamiltonian. For three-flavors, there can be two Majorana phases. Several two-flavor simulations were run with Majorana phases different from zero. It was found, given our chosen initial conditions and parameters, that the Majorana phase affected the spin transformations only negligibly. There are perhaps two reasons that the Majorana phase would not significantly affect our spin transformations. First, we note that equation \ref{eqn:massmtx} can be multiplied out and rewritten, with $c_\theta\equiv\cos\theta_\text{V}$ and $s_\theta\equiv\sin\theta_\text{V}$ for brevity, as \cite{2015PhLB..747...27C,2015PhRvD..91l5020K}:
\begin{widetext}
\begin{eqnarray}
m=\frac{m_1+m_2}{2}
\begin{pmatrix}c_\theta^2+e^{-i\alpha}s_\theta^2 && (e^{-i\alpha}-1)s_\theta c_\theta \\
(e^{-i\alpha}-1)s_\theta c_\theta && s_\theta^2+e^{-i\alpha}c_\theta^2
\end{pmatrix}
+\frac{m_2^2-m_1^2}{2(m_1+m_2)}
\begin{pmatrix}e^{-i\alpha}s_\theta^2-c_\theta^2 && (e^{-i\alpha}+1)s_\theta  c_\theta \\
(e^{-i\alpha}+1)s_\theta c_\theta && e^{-i\alpha}c_\theta^2-s_\theta^2
\end{pmatrix} .
\label{eqn:massmtxmult}
\end{eqnarray}
\end{widetext}
Notice that if $m_1+m_2 \gg m_2-m_1$, as was the case in all of our simulations, the first term in equation \ref{eqn:massmtxmult} will dominate. Furthermore, if the spin transformations occur before significant flavor transformations, then for the $\nu_e\rightarrow\bar{\nu}_e$ transformation channel it is really only the top left term in the mass matrix (the $\frac{m_1+m_2}{2}(c_\theta^2+e^{-i\alpha}s_\theta^2)$ term) that matters. This is because $H^{sf}_{\nu\nu}$ will be diagonal if the neutrinos are all in flavor eigenstates and so $(H^{sf})_{11}$ will only have contributions from this one mass matrix term. The first reason the Majorana phase only negligibly affected our simulations is that for the above-mentioned term in the mass matrix the Majorana phase, $e^{-i\alpha}$, multiplies a $\sin^2\theta_\text{V}$. Since we take $\theta_\text{V}=8.7^\circ$ in our analysis, the relevant mass matrix term involving the Majorana phase will be very small: $\sin^2\theta_\text{V} \approx .023$ vs $\cos^2\theta_\text{V} \approx .98$. As a result, the Majorana phase, even if set to $\pi$, cannot significantly affect the pertinent term in the mass matrix for the given parameters used in our simulations.

The second potential reason the Majorana phase did not affect our simulations significantly is that the Majorana phase can only affect the adiabaticity parameter and not the spin resonance condition itself. The Majorana phase does not produce a vacuum splitting in the energy between neutrinos and antineutrinos, i.e. it only appears in $H^{sf}$, not in $H_{vac}$, $H_{m}$ or $H_{\nu\nu}$. The spin transformations are really set by nonlinear effects keeping the neutrinos near resonance. Nonlinear feedback keeping spin transformation adiabatic and \lq\lq tracking\rq\rq\ over a range of densities is a key feature of three-flavor, two-flavor and one-flavor calculations \cite{2014arXiv1406.6724V}. The ratio of electron neutrinos to antineutrinos is set by the fact that the neutrino-neutrino term in the diagonal Hamiltonian, $H_{11}$, has to cancel out the matter term in order for there to be such tracking. The matter potential that we used was the same for different simulations; therefore, as long as the neutrinos were made to track $H_{11} \approx 0$ over the same physical interval, the neutrino to antineutrino ratio had to remain roughly the same no matter what the Majorana phase was. Our simulations show that the slight change in adiabaticity introduced by changing the Majorana phase was not enough to significantly affect the tracking behavior, and therefore did not affect the spin transformations. 

\subsection{Onset of Transformations}\label{sec:onsetoftransformations}

An important aspect of spin and flavor transformations that we have been able to probe with our code is the locations, relative and absolute, where these spin and flavor transformations are most pronounced. Due to the nature of the spin transformation's resonance conditions, spin transformations have so far been found to occur only in locations where $Y_e\lesssim 1/3$. In addition, simulations in which the electron fraction profile was set to $Y_e\approx 1/3$ far from the neutrino sphere, all else being the same, were not able to produce significant spin transformations. This is simply due to the fact that as we get farther from the neutrino sphere the geometric dilution will necessarily dilute the spin-flip Hamiltonian $H^{sf}$. If the spin-flip Hamiltonian is too small, then, by equation \ref{eq:adiabat} spin transformations will be highly nonadiabatic at resonance and it will be difficult for significant spin transformations to set in. 

Significant flavor transformations in our simulations have so far always occurred after spin transformations have ceased (see figures \ref{fig:solarresult}, \ref{fig:atmresult}, and \ref{fig:3fresult} ). As the flavor transformations do depend on the neutrino spin content, i.e. on the ratio of neutrinos to antineutrinos (see equations \ref{eqn:hvv},\ref{eqn:lambdamatrix}) the spin transformations have the potential to affect flavor transformations. Vice versa, if flavor transformations were to occur prior to the onset of spin transformations, it is also possible that spin transformations could be affected. However, as the spin transformations appear to require an extremely large neutrino flux to be significant - the vacuum Hamiltonian does not contribute to spin-flip - it appears that significant spin transformations, if they do occur, are likely to occur closer to the neutrino sphere than flavor transformations.

\subsection{Flavor Transformations}\label{sec:flavorT}         

As was discussed earlier, flavor transformations have not been able to feed back into spin transformations in any way in our simulations since spin transformations begin and end before flavor transformations even start. In the solar mass-squared splitting case, spin transformations did not qualitatively change the flavor transformations. However, for the atmospheric mass-squared splitting and for three-flavor simulations, the spin transformations did significantly impact the flavor transformations. The process by which spin transformations change flavor transformations is simply through the production of antineutrinos. From equations \ref{eqn:hvv} and \ref{eqn:lambdamatrix} we can see that a flux of antineutrinos affects the energy splitting between the two-flavor states. 

The simulation with no spin coherence and pure flavor transformations essentially reproduced previous results found in \cite{Duan08} and \cite{Cherry:2010lr}. Due to the extreme neutrino fluxes found during the neutronization burst, the neutrinos all go through a lepton number ($n_{\nu_e}-n_{\bar{\nu}_e}\approx n_{\nu_e}$, very few antineutrinos are present) nonconserving flavor resonance together in a \lq\lq neutrino background assisted MSW-like resonance\rq\rq. Basically, at the radius where a representative energy neutrino would go through the MSW resonance, the neutrino self-coupling locks neutrinos of all energies together so that all neutrinos go through the resonance together. This produces a neutrino spectrum which is overwhelmingly in the $x$-neutrino state. Immediately after going through this MSW-like resonance, the neutrinos are locked into collective oscillations and finally when those collective oscillations die out a lepton number conserving swap is formed at energy $E_l\approx 20 \, \text{MeV}$. Note that the conserved lepton number is actually the \lq\lq mass basis lepton number\rq\rq\ which would be $L=(n_{\nu_1}-n_{\bar{\nu}_1})-(n_{\nu_2}-n_{\bar{\nu}_2})$ where $\nu_1$ and $\nu_2$ are the vacuum mass eigenstates. However, we used a small mixing angle for two-flavor simulations so that the flavor lepton number is approximately conserved. For qualitative discussion, we need not make the distinction. The swap energy $E_l$ is determined by a conservation of the lepton number immediately after the MSW-like resonance (mostly $x$-neutrinos). The swap energy is therefore high because the MSW-like resonance was quite efficient at destroying $n_{\nu_e}$ \cite{movies}. 

The simulation with spin coherence was qualitatively and quantitatively different. Due to the presence of antineutrinos, the flavor transformations are able to undergo a classic spectral swap without first undergoing the neutrino background assisted MSW-like resonance. Also, due to the presence of a large number of antineutrinos, the MSW-like resonance may not be nearly as strong as for when there are no antineutrinos because the neutrino-neutrino Hamiltonian is suppressed by the presence of antineutrinos. Indeed, if $n_{\nu_e}=n_{\bar{\nu}_e}$, then we can see from equations \ref{eqn:hvv} and \ref{eqn:lambdamatrix} that $H_{\nu\nu}=0$. The neutrinos and antineutrinos alike are locked into collective oscillation modes, but the spectral swap that develops conserves a lepton number which was not significantly affected by the MSW-like resonance. As such, the swap energy $E_l\approx 9 \, \text{MeV}$ was much lower because the MSW-like resonance was not able to convert the vast majority of electron neutrinos into the $x$-neutrino state \cite{movies}.

\section{Conclusion}\label{sec:conclusion}

This paper presents the first multiflavor simulations of coherent neutrino spin transformations using a neutrino bulb geometry. We explored a variety of initial conditions and parameters using an O-Ne-Mg supernova density profile and have presented results for those initial conditions and parameters which produced significant spin transformation. We found that it is likely, given the nature of the spin-flip Hamiltonian, that the spin transformations, if they occur, would occur prior to the onset of significant flavor transformations. As a result, the spin transformations (more precisely, the neutrino to antineutrino ratio produced by spin transformations) are not affected by the flavor structure of neutrinos (i.e., mass splittings, mixing angles, and number of flavors). However, there is potential for spin coherence to change the nature of the subsequent flavor transformations.

Our simulations found that, for significant spin transformations to occur, an unrealistically massive neutrino ($m_\nu = 10\,\text{eV})$, a large neutrino luminosity, and an electron fraction profile which hovered near $Y_e \lesssim 1/3$ for several hundred kilometers were required. For the parameters and initial conditions considered in our two-flavor simulations, a Majorana phase did not significantly alter the spin transformations. Changing the neutrino-to-antineutrino ratio requires changing the so-called tracking behavior ($(H_m)_{11}+(H_{\nu\nu})_{11}\approx 0$) of the neutrino Hamiltonian, which a change in the Majorana phase fails to affect.

Geometric dilution of the neutrinos, and the steep density dropoff combine to make tracking difficult, even with the extreme neutrino fluxes encountered during the neutronization burst. However, our simulations have not yet included other potential feedback loops such as the $Y_e$ feedback \cite{2014arXiv1406.6724V,McLaughlin:1999fk,Fetter:2003lr}. Perhaps the inclusion of other feedback mechanisms could enable spin transformations to occur with more realistic values of neutrino rest masses. If such feedback mechanisms help initiate tracking, then, as we have shown, a significant spin transformation in the neutrino population could significantly and qualitatively change the subsequent flavor evolution of these neutrinos. Unless it can be proven that no such mechanism exists in the supernova environment, which would make the spin coherence resonance adiabatic enough to engage the tracking behavior, spin degrees of freedom would necessarily have to be considered when considering neutrino flavor transformations. As neutrinos of different flavors interact with matter differently, changing the neutrino content could lead to ramifications on the nucleosynthesis ($r$-process) of elements during core collapse supernovae \cite{2016AIPC.1743d0001B,2016EPJWC.10906005W} and the reheating of the initial supernova shock \cite{Bethe:1985lr,2006A&A...447.1049B,2013RvMP...85..245B}.

\begin{acknowledgements}
We would like to thank J.~Carlson, J.F.~Cherry, V.~Cirigliano, L.~Johns, C.~Kishimoto, J.T.~Li, S.~Tawa, and A.~Vlasenko for valuable conversations. This work was supported in part by NSF Grants No. PHY-1307372 and No. PHY-1614864 at University of California, San Diego. We also acknowledge a grant from the University of California Office of the President.
\end{acknowledgements}
\bibliography{allref}

\end{document}